# Real-Time Prediction for Athletes' Psychological States Using BERT-XGBoost: Enhancing Human-Computer Interaction

Chenming Duan[1,5], Zhitao Shu[2,6], Jingsi Zhang [3,7], Feng Xue [4,8,*]

[1] Russian Sport University, Moscow, Russia
[2] Vanderbilt University, Nashville, TN, USA
[3] Khoury College of Computer Sciences, Northeastern University, Boston, MA, USA
[4] Northwest University For Nationalities,  Lanzhou City, Gansu Province, China

[5] 790212967@qq.com
[6] zhitao.shu@vanderbilt.edu
[7] zhang.jings@northeastern.edu
[8] 264626150@qq.com
*corresponding author

**Abstract.** Understanding and predicting athletes' mental states is crucial for optimizing sports performance. This study introduces a hybrid BERT-XGBoost model to analyze psychological factors such as emotions, anxiety, and stress, and predict their impact on performance. By combining BERT's bidirectional contextual learning with XGBoost's classification efficiency, the model achieves high accuracy (94%) in identifying psychological patterns from both structured and unstructured data, including self-reports and observational data tagged with categories like emotional balance and stress. The model also incorporates real-time monitoring and feedback mechanisms to provide personalized interventions based on athletes' psychological states. Designed to engage athletes intuitively, the system adapts its feedback dynamically to promote emotional well-being and performance enhancement. By analyzing emotional trajectories in real-time, it offers empathetic, proactive interactions. This approach optimizes performance outcomes and ensures continuous monitoring of mental health, improving human-computer interaction and providing an adaptive, user-centered model for psychological support in sports.

## 1.Introduction

In competitive sports, athletes' psychological states are pivotal to their performance. Factors such as emotions, anxiety, and stress significantly impact not only individual outcomes but also team dynamics and long-term career trajectories. Recognizing and addressing these psychological elements is essential for both athletes and sports organizations.

Traditional psychological assessments, like surveys and interviews, often fail to capture the complexity of athletes' mental states. However, advancements in machine learning, natural language processing (NLP), and human-computer interaction (HCI) now enable the analysis of psychological patterns from diverse data sources, including self-reports and physiological metrics. HCI plays a crucial role by facilitating interactive systems, such as touchscreen or voice-activated interfaces, where athletes can input their emotions more naturally. These systems can adapt to emotional cues, offering personalized feedback, like calming visuals or music when stress is detected, enhancing the user experience.

This study proposes a hybrid BERT-XGBoost model to predict athletes' psychological states and develop personalized intervention strategies. BERT excels at extracting rich contextual information from textual data, while XGBoost provides efficient and interpretable classification. By combining both, the model offers a powerful framework for analyzing psychological factors affecting performance. The integration with HCI ensures that the system can adapt to real-time emotional cues, such as changes in facial expressions or tone of voice, optimizing the feedback and intervention provided to athletes.

The dataset includes psychological metrics and self-reports tagged with emotional states like anxiety, stress, and performance factors. The study identifies key predictors of performance-limiting psychological states and evaluates the hybrid model's accuracy. It also explores how adaptive feedback, central to HCI, enhances athletes' engagement and trust in the system.

The significance of this research lies in its dual contribution: enhancing the understanding of athletes' mental states and advancing HCI-driven intervention strategies that optimize human-computer interaction through emotional awareness. The outcomes are expected to inform sports psychologists, coaches, and athletes in real-time decision-making and long-term psychological support strategies, offering an innovative path toward personalized mental health solutions in competitive sports.

## 2. Literature Review

Sentiment analysis has emerged as a critical area of research, leveraging advanced machine learning techniques to analyze textual data for emotional insights. The integration of BERT (Bidirectional Encoder Representations from Transformers) and other machine learning models has significantly advanced the field, particularly in aspect-based sentiment analysis and related tasks.

Hoang et al. demonstrated the potential of BERT in aspect-based sentiment analysis, highlighting its ability to capture contextual semantics and fine-grained relationships between words in a sentence. Their study underscored the importance of pre-trained language models in outperforming traditional methods in sentiment classification tasks **[1]**. Similarly, Xu et al. extended the capabilities of BERT through post-training techniques, which improved its performance in review reading comprehension and aspect-based sentiment analysis by tailoring the model to domain-specific datasets **[2]**.

Li et al. explored the end-to-end application of BERT for aspect-based sentiment analysis, showcasing how its representation learning abilities streamline the process of sentiment categorization without requiring additional feature engineering. Their findings emphasize the model's robustness in extracting sentiment from diverse textual inputs **[3]**.

Singh et al. applied BERT to analyze the impact of the COVID-19 pandemic on social life through sentiment analysis of social media data. This work demonstrated BERT's efficacy in understanding public sentiment during crises, emphasizing the model's adaptability to dynamic and event-specific data **[4]**.

While BERT excels in representation learning, its integration with machine learning algorithms like XGBoost has further enhanced its performance in sentiment analysis. Samih et al. proposed combining improved word embeddings with XGBoost for enhanced sentiment classification. Their approach illustrated how ensemble learning methods can leverage semantic representations to improve predictive accuracy **[5]**.

Hama Aziz and Dimililer introduced SentiXGBoost, an ensemble classifier that combines BERT's contextual embeddings with XGBoost's efficient decision tree learning. Their experiments on social media datasets revealed the hybrid model's superior performance in handling large-scale, noisy textual data **[6]**. Chandrasekaran et al. extended this concept to multimodal sentiment analysis, incorporating both textual and non-textual features into a deep neural network architecture enhanced by XGBoost. Their findings highlighted the hybrid model's adaptability to complex data formats and its potential in diverse application areas **[7]**.

Alaparthi and Mishra reviewed the evolution of sentiment analysis techniques, focusing on BERT's trajectory as a dominant model in this domain. Their work provided a comprehensive comparison of BERT with other approaches, reinforcing the advantages of its bidirectional context representation and pre-trained architecture **[8]**.

The literature consistently highlights the effectiveness of BERT in extracting deep semantic features and the complementary strengths of XGBoost in classification tasks. The integration of these models addresses the challenges of noise, scalability, and feature nonlinearity in sentiment analysis, setting a new benchmark for performance in the field.

**3. Data**

*3.1 Data introduction*

This study uses a dataset derived from self-reports, observational records, and performance logs of athletes, collected during training and competitions. The data includes narrative descriptions of athletes' mental states and emotion tags such as Anxiety, Stress, Burnout, and Focus, annotated by sports psychologists for accuracy. The dataset captures a range of psychological responses in competitive sports, providing valuable insights for model development. By combining textual analysis with context, it supports the creation of advanced machine learning models like the BERT-XGBoost hybrid, which can predict psychological states and improve athlete intervention strategies.

*3.2 Descriptive statistical analysis*

The statistical analysis of the distribution of mental health status found that the sample showed a diverse distribution of mental health characteristics. The data showed that the normal group accounted for 31.0% of the total, constituting the largest proportion; followed by depression symptoms (29.2%); suicidal (Suicidal) ranked third, accounting for 20.2%, which deserves high attention. Anxiety symptoms accounted for 7.3%, bipolar disorder (5.3%), stress (Stress) related problems accounted for 4.9%, and personality disorder (Personality disorder) accounted for 2.0%.

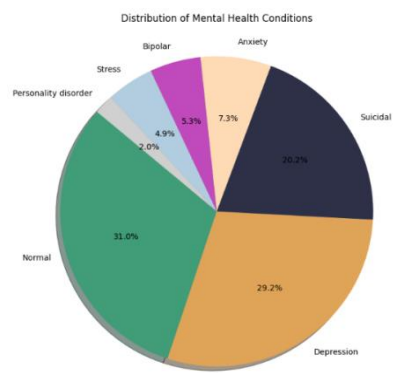

**Figure 1.** Distribution of Mental Health Conditions

The data reveals key insights: nearly 70% of samples exhibited mental health issues, emphasizing the need for intervention; depression and suicidal tendencies comprised 50% of severe cases, underlining the importance of preventive measures; and the variety of mental health issues calls for personalized intervention strategies. To analyze text features, the study conducted multi-dimensional statistical extraction, calculating key indicators: (1) text length (statement_length), (2) word count (num_words), (3) average word length (avg_word_length), and (4) vocabulary size (vocabulary_size). These metrics provide a quantitative basis for analyzing linguistic features, emotional patterns, and the author's expression habits, forming the foundation for further analysis.

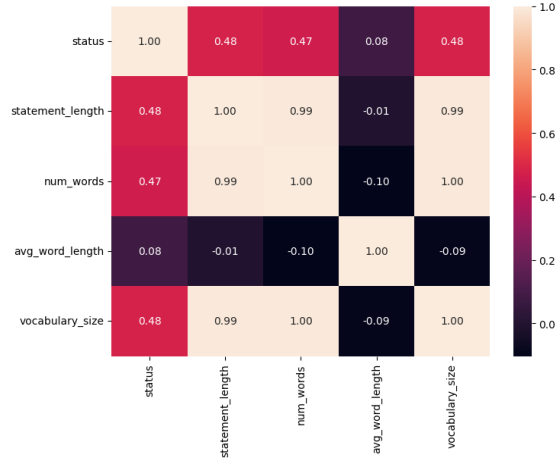

**Figure 2.** Correlation heat map

Heat map analysis shows that there is a significant positive correlation between text length, word number and vocabulary, indicating that these characteristics play an important role in measuring text complexity and richness. At the same time, the average word length has a low correlation with other characteristics, reflecting its independence in lexical complexity analysis. These characteristics provide an important quantitative basis for subsequent research on language features and emotional expression patterns.

## 4. Model

*4.1 Model introduction*
BERT is a pre-trained deep learning model based on the Transformer architecture, designed to understand the context of words in text by capturing bidirectional relationships. Unlike traditional models, which read text either left-to-right or right-to-left, BERT processes text in both directions simultaneously, enabling it to understand the full context of a word within a sentence.

BERT leverages the Transformer's self-attention mechanism, which assigns weights to each word in a sequence relative to other words, enabling the model to focus on contextually significant terms. Mathematically, the self-attention scores are calculated as:

$$Attention(Q, K, V) = softmax(QK^T / sqrt(d_k))V$$

BERT uses multiple Transformer layers, each with attention heads and feedforward neural networks, to capture abstract representations of input data. Fine-tuned for specific tasks, BERT adds a task-specific output layer, such as a classifier for text classification. XGBoost, a gradient-boosted decision tree (GBDT) algorithm, excels in large-scale data and high-dimensional features. It improves predictive power by iteratively adding decision trees to fit residuals, using regularization to prevent overfitting and selecting optimal split points based on gain calculations.

*4.2 Model processing process*
In this study, the BERT-XGBoost hybrid model architecture is used for text classification tasks. In the process of model construction, BERT (Bidirectional Encoder Representations from Transformers) acts as a feature extractor to transform the original text into a high-dimensional semantic representation through a pre-trained language model. Specifically, the BERT model first performs operations such as text standardization, word segmentation and special tag addition through the pre-processing layer, and then uses a 12-layer transformer encoder to generate a 768-dimensional dense representation. These vectors contain deep semantic information of the text, effectively capturing features at multiple levels such as word order relationship, syntactic structure and context semantics.

The model uses BERT-generated 768-dimensional feature vectors as input to the XGBoost classifier. XGBoost employs ensemble learning to build tree models, capturing complex nonlinear relationships in high-dimensional feature spaces. The system uses 500 decision trees, an early stopping strategy (early_stopping_rounds = 10), and a learning rate of 0.05 to prevent overfitting and ensure stable convergence. This hybrid approach, combining BERT's representation learning with XGBoost's classification efficiency, offers an effective, computationally efficient solution for text classification tasks.

*4.3 Model results*

```
              precision    recall  f1-score   support

           0       0.95      0.92      0.93      1260
           1       0.88      0.82      0.85      1220
           2       0.88      0.88      0.88      1187
           3       0.97      0.99      0.98      1252
           4       0.97      1.00      0.98      1215
           5       0.96      0.99      0.98      1210
           6       1.00      1.00      1.00      1252

    accuracy                           0.94      8596
   macro avg       0.94      0.94      0.94      8596
weighted avg       0.94      0.94      0.94      8596
```

**Figure 3.** Model classification result

From the experimental results, it can be seen that the BERT-XGBoost hybrid model exhibits excellent performance in text classification tasks. The overall accuracy rate reaches 0.94, indicating that the model can accurately classify text in most cases. The F1 scores for all categories are generally high, especially for categories 3, 4, 5, and 6. The F1 scores are close to or reach 1.00, indicating that the model has very strong classification ability on these categories. This efficient classification performance reflects the robustness and accuracy of the model when processing complex text data.

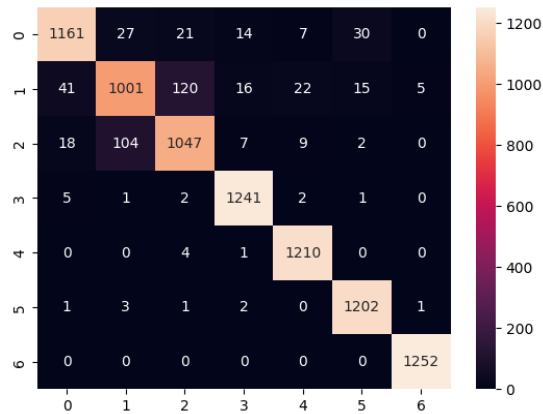

**Figure 4.** Confusion matrix thermography

The confusion matrix analysis reveals frequent misclassifications in categories 0, 1, and 2, particularly between categories 1 and 2, likely due to their feature similarity. However, category 6 showed perfect classification, indicating strong model distinction. This suggests the need for more targeted feature extraction and model tuning for specific categories. BERT's ability to capture deep semantic information provides rich input for XGBoost, enhancing feature representation. The model optimization, with 500 decision trees, early stopping, and a learning rate of 0.05, prevents overfitting and ensures stable convergence, improving generalization and classification reliability. The combined BERT-XGBoost approach offers an efficient solution for text classification tasks.

**5. Conclusions and Suggestions**

This study demonstrates the effectiveness of a hybrid BERT-XGBoost model for predicting athletes' psychological states and understanding their impact on performance. The model integrates BERT's contextual learning and XGBoost's efficient classification capabilities, achieving high accuracy in predicting emotions, stress levels, and anxiety. The results highlight the model's ability to handle both structured and unstructured data, revealing psychological trends that influence athletic performance. Anxiety and stress were found to significantly correlate with performance fluctuations, underscoring the importance of early detection and management.

Despite the model's success in capturing deep semantic features and non-linear relationships, challenges remain in differentiating closely related psychological states, such as anxiety and stress. Future work should focus on incorporating additional behavioral, physiological, and contextual features to improve classification precision. Real-time emotional feedback, combined with biometric sensors (e.g., heart rate variability, facial expressions), could enhance the model's adaptability and provide more accurate predictions.

Deploying this model as part of a real-time monitoring system would enable coaches and sports psychologists to identify athletes at risk, offering immediate interventions based on the model's predictions. Integrating human-computer interaction (HCI) technologies would facilitate intuitive feedback loops, allowing athletes to engage with the system and receive tailored suggestions for managing their mental states.

Additionally, future studies should consider integrating multimodal data sources, such as video analysis and voice tone, and expanding the dataset to include diverse athletes across sports, competition levels, and cultural contexts. This will improve the model's generalizability and accessibility, making it a valuable tool for enhancing athletes' psychological well-being and performance. By fostering a deeper human-computer partnership, the model can empower athletes to take control of their mental health and optimize their performance.